\documentclass[12pt,a4paper]{article}
\usepackage[margin=27mm]{geometry}
\usepackage[T1]{fontenc}
\usepackage{newtxtext}
\usepackage{amsmath,amsthm,mathtools}
\usepackage{newtxmath}
\usepackage{booktabs,float}
\usepackage{tikz}
\usepackage[authoryear,round]{natbib}
\usepackage[hidelinks]{hyperref}
\usepackage[nopatch=footnote]{microtype}
\usepackage{setspace,needspace}
\numberwithin{equation}{section}
\newtheorem{theorem}{Theorem}[section]
\newtheorem{proposition}[theorem]{Proposition}
\newtheorem{corollary}[theorem]{Corollary}
\newcommand{\R}{\mathcal R}
\newcommand{\WPS}{\operatorname{WPS}}
\newcommand{\SES}{\operatorname{SES}}
\newcommand{\Sh}{\operatorname{Sh}}
\newcommand{\ind}{\mathbf 1}
\title{\bfseries Weighted Proportional Toll Sharing on General Highway Networks}
\author{Erfang Shan \and Yan Chu}
\date{\small School of Management, Shanghai University, Shanghai 200444, China\\
\texttt{efshan@shu.edu.cn} (E. Shan)}
\begin{document}
\maketitle

\begin{abstract}
We extend the Segments Equal Sharing (SES) method of \citet{wu2024} from line highways to general highway networks, including networks with cycles. The input is an exogenously specified profile of route flows. We introduce the Weighted Proportional Sharing (WPS) method, which distributes each route's mileage revenue in proportion to weighted segment lengths and its fixed-surcharge revenue in proportion to positive segment surcharges. Route SES is recovered under homogeneous weights within each revenue component. We show that WPS is the sum of two weighted Shapley allocations and that the underlying route games extend the dividend and Shapley representation of linear SES. Two axiomatic characterizations identify WPS. The first uses additivity and proportional treatment of segments used on every positive-revenue route. The second combines sub-highway efficiency and weighted average balanced contributions under a route cancellation rule. We also extend Segments Proportional Sharing to general networks, identify it with the $\tau$ value of the corresponding route game, and characterize it axiomatically. For Segments Compensated Sharing we characterize a weighted average tree allocation by consistency across rooted branch settlements and averaging under a specified root lottery. A cyclic example shows that different route splits can change the allocation even when origin--destination volumes and aggregate toll revenue are held fixed.
\end{abstract}


\section{Introduction}

An integrated tolling system may collect a single payment for a trip that crosses segments operated by several concessionaires. A settlement rule specifies how the collected revenue is attributed to the participating segments. This is a revenue allocation problem among segments, distinct from the problem of setting tolls or allocating infrastructure costs among users. The distinction matters when the total payment is pooled and the agreement among operators determines which revenues they can claim.

\subsection{Related literature and contribution}

Highway cost allocation and highway toll settlement assign money to different agents. In the former, the users pay toward the cost of making the road available. \citet{kuipers2013} define highway cost games for users who require consecutive sections of a common road and develop formulas for the Shapley value and nucleolus. \citet{sudholter2017} characterize cost-sharing methods for highway problems, and \citet{liweighted2026} introduce weights for heterogeneous users in highway construction cost sharing. In our model the customers' payments have already been collected. The players are road segments, and the two weights of a segment describe its eligibility and relative share in the mileage and surcharge components.

Work on other transport systems illustrates why the player set and the characteristic function must be chosen together. \citet{rosenthal2017} studies the allocation of rapid-transit costs among users travelling on shortest paths in a general network. \citet{estan2021} studies how cities served by a tram line can share a fixed system cost, including proportional and equal division principles. These applications establish precedents for using network information and proportionality in transport cost allocation. The settlement problem here instead starts from the realised or supplied use of each highway route and asks how that route's collected revenue is credited to the segments.

\citet{wu2024} study this problem on a line highway. Their first rule, the Segments Equal Sharing (SES) method, distributes the toll associated with each trip equally among the segments it uses. They characterize SES by additivity, efficiency, an inessential segment property and segment symmetry, and identify it with the Shapley value of a segments allocation game. Their Exits Equal Sharing and Entrances Equal Sharing methods address settlements based on exit and entrance tolls, respectively. \citet{soto2026} further study the linear setting and introduce Segments Proportional Sharing and Segments Compensated Sharing, with representations involving the $\tau$-value and the average tree solution.

The cooperative-game analysis draws on graph-restricted games \citep{myerson1977,owen1986} and weighted Shapley values \citep{kalai1987}. A graph restriction uses communication links to determine which groups can cooperate. In our model the route data specify directly which sets of segments generate the revenue being divided. This distinction matters when a graph has cycles. For a physical route that is disabled, an alternative connection elsewhere in the communication graph need not restore any of that route's revenue. It also matters for fixed surcharges, because two eligible segments can lie on the same route even if a zero-surcharge segment separates them.

\subsection{Approach and organization}

Two features of general highway networks motivate our extension of SES. First, a pair of interchanges can be connected by several routes. An origin--destination total may therefore be compatible with several different patterns of segment use. Second, the toll paid on a route can combine a mileage charge and fixed surcharges associated with particular segments. Equal division of the combined payment treats these two charging bases identically. A weighted rule can instead distinguish the length and attributes of the segments that support mileage charging from the facilities for which a fixed surcharge is collected.

We model each route and its class-specific flow as an input. Our Weighted Proportional Sharing (WPS) method splits each route's mileage revenue by weighted length and its surcharge revenue by the applicable positive surcharges. This extends SES in two directions. Equal sharing can be applied route by route on an arbitrary network, and component-specific weights allow that rule to distinguish segments with heterogeneous attributes. On a line, route SES agrees exactly with the rule of \citet{wu2024} when interval toll revenues are matched. Their path-dividend formula and Shapley representation are special cases of the results below. Our first axiomatic characterization also recovers their characterization when specialized to the line and equal weights.

The second characterization uses the allocation effects of cancelling specified routes that use a segment. It gives a fairness interpretation in terms of reciprocal impacts, averaged over a group of segments that shares its route revenues internally. The operation holds all unaffected route flows fixed, so it can be used on a network with cycles without modelling an equilibrium response. The resulting axioms describe alternative settlement principles for a given flow profile. Their implications concern the resulting allocation within the model.

Section~\ref{sec:prelim} introduces TU games, coalition dividends, Shapley values and graph restriction. Section~\ref{sec:model} sets out the model and allocation rules. Section~\ref{sec:game} gives the cooperative-game representation. Section~\ref{sec:axioms} states the two WPS characterizations. Section~\ref{sec:sps-scs} extends SPS and SCS. It characterizes SPS and provides a recursive characterization of a weighted average tree rule for SCS. Section~\ref{sec:example} illustrates the effect of a route split on a cyclic network, and Section~\ref{sec:conclusion} discusses the scope for further extensions. Proofs are collected in Appendix~\ref{app:proofs}. Appendix~\ref{app:independence} tests logical independence for the WPS, SPS and SCS characterizations.

\section{Cooperative game preliminaries}\label{sec:prelim}

\subsection{Transferable utility games and coalition dividends}

A transferable utility (TU) game is a pair $(N,v)$, where $N$ is a finite player set and $v:2^N\to\mathbb R$ is a characteristic function with $v(\varnothing)=0$. A subset $S\subseteq N$ is a coalition and $v(S)$ is the total worth available to its members. A payoff vector $x\in\mathbb R^N$ is efficient if $\sum_{i\in N}x_i=v(N)$. A value associates a payoff vector with each game on its domain. In our application players are highway segments. Coalition worth is measured in toll revenue, so amounts assigned to different segments can be added and compared.

For each nonempty $A\subseteq N$, the unanimity game $u_A$ is defined by $u_A(S)=\ind\{A\subseteq S\}$. Thus $u_A$ generates one unit of worth precisely when all members of $A$ participate. The unanimity games form a basis of the vector space of TU games. The Harsanyi dividend $\Delta_v(A)$ of $A$ is the coefficient of $u_A$ in the expansion of $v$ \citep{harsanyi1959}.

\Needspace{7\baselineskip}
With $\Delta_v(\varnothing)=0$, the expansion and its inverse are
\begin{align}
v(S)&=\sum_{A\subseteq S}\Delta_v(A),\label{eq:dividend-expansion}\\
\Delta_v(A)&=\sum_{B\subseteq A}(-1)^{|A|-|B|}v(B).\label{eq:mobius}
\end{align}
Equivalently, $\Delta_v(A)$ is the worth of $A$ less the dividends of all its proper subcoalitions. Coalition worth and coalition dividend can therefore differ. A group of segments may receive substantial route revenue even when it has zero dividend, because the revenue is already recorded in dividends of smaller groups.

For example, let $N=\{1,2,3\}$ and $v=60u_{\{1,2\}}+30u_{\{2,3\}}$. Coalition $\{1,2,3\}$ has worth 90 but dividend zero. Its entire worth is accounted for by the dividends 60 and 30 of the two pairs. Singleton coalitions have no worth in this example. This distinction is important on a road network with a cycle, where a coalition consisting of all the segments in the cycle may contain several revenue-generating routes.

\subsection{The Shapley value and positive weights}

The Shapley value averages each player's marginal contribution over the possible orders in which players can join the coalition \citep{shapley1953}. For $i\in N$ it is
\begin{equation}\label{eq:shapley-marginal}
\Sh_i(v)=
\sum_{S\subseteq N\setminus\{i\}}
\frac{|S|!\,(|N|-|S|-1)!}{|N|!}
\bigl[v(S\cup\{i\})-v(S)\bigr].
\end{equation}
In a unanimity game $u_A$, a player can make a nonzero contribution only if it is the last member of $A$ to arrive. Each member of $A$ has the same probability of doing so. Hence, by linearity and \eqref{eq:dividend-expansion},
\begin{equation}\label{eq:shapley-dividend}
\Sh_i(v)=\sum_{\substack{A\subseteq N\\ i\in A}}\frac{\Delta_v(A)}{|A|}.
\end{equation}

For a positive weight vector $w\in\mathbb R_{++}^{N}$, the weighted Shapley value considered here divides each dividend among the members of its coalition in proportion to their weights \citep{kalai1987}. Writing $w(A)=\sum_{j\in A}w_j$, we have
\begin{equation}\label{eq:weighted-dividend}
\Sh_i^w(v)=\sum_{\substack{A\subseteq N\\i\in A}}\frac{w_i}{w(A)}\Delta_v(A).
\end{equation}
This rule is linear in $v$ for fixed weights and efficient, since the shares of each dividend sum to one. Equal weights give \eqref{eq:shapley-dividend}. A common rescaling of all weights leaves the allocation unchanged. For the three-player example above, the ordinary Shapley value is $(30,45,15)$. With $w=(1,2,1)$ the weighted value is $(20,60,10)$, obtained by dividing the first dividend in the ratio $1:2$ and the second in the ratio $2:1$.

\citet{kalai1987} also allow ordered priority classes in a weight system. We use the positive-weight form \eqref{eq:weighted-dividend}. For a revenue component with zero-weight segments, the game for that component is defined on its positive-weight player set, and the associated payoff vector is extended by zero. This choice keeps all denominators in \eqref{eq:weighted-dividend} positive and identifies explicitly which segments are entitled to share that component.

\subsection{Communication graphs and connected coalitions}

A communication graph $H=(N,L)$ has players as vertices and bilateral links as edges. For $S\subseteq N$, let $\mathcal C_H(S)$ denote the connected components of the graph induced by $S$. The Myerson graph restriction of a TU game is
\begin{equation}\label{eq:myerson}
v^H(S)=\sum_{C\in\mathcal C_H(S)}v(C).
\end{equation}
The Myerson value is the Shapley value of $v^H$ \citep{myerson1977}. A game $v$ is component additive on $H$ if $v^H=v$. The dividends of a graph-restricted game vanish on disconnected coalitions \citep{owen1986}. This does not require the communication graph to be a tree. The equivalent condition below is useful for interpreting the road games.

\begin{proposition}[Component additivity]\label{prop:components}
A TU game is component additive on $H$ if and only if $\Delta_v(A)=0$ for every disconnected nonempty coalition $A$ in $H$.
\end{proposition}

There are two graphs in the highway application. The physical network $G=(V,E)$ has road segments as edges. The line graph $L(G)$ has the segments as vertices, and two vertices of $L(G)$ are adjacent when the corresponding segments share a physical endpoint. A segment coalition $F\subseteq E$ is connected in $L(G)$ exactly when the physical subgraph formed by its edges and their endpoints is connected. In particular, a simple route in $G$ has a connected segment set in $L(G)$. Even a physical tree can have a cyclic line graph. For example, three roads incident to the same junction give a triangle in $L(G)$.

Connectivity alone does not identify the dividend supports of a route game. A branch or a cycle may be connected even though its full segment set cannot be traversed as one simple route. Conversely, projecting a simple route onto its positive-surcharge segments can leave a disconnected set in the induced subgraph of $L(G)$. These observations distinguish the coalition structure of the full-route game from that of the projected surcharge game.

\section{Network model and allocation rules}\label{sec:model}

\subsection{Routes and revenues}

Let $G=(V,E)$ be a finite connected undirected highway network. Nodes represent entrances, exits or interchanges, and edges represent the segments among which toll revenue is allocated. Cycles and branching are allowed. Fix a finite family $\R$ of admissible nonempty simple routes. For $r\in\R$, write $P_r\subseteq E$ for its segment set. Different routes for the same origin--destination pair can have positive flows. Direction-specific records may be kept separately whenever needed. The rule sums their contributions whenever they use the same segments. We allow zero flows so that every route cancellation leaves the same network and route family.

Let $q_r^{(i)}\ge 0$ be the flow of vehicles of class $i$ on route $r$, and let $c_i\ge 0$ be the class-specific mileage rate. For a segment $e$, let $l_e>0$ denote its length, $\alpha_e\ge 1$ its mileage multiplier, and $\beta_e\ge 0$ its per-vehicle fixed surcharge. A multiplier can reflect a tariff adjustment for a segment attribute. These parameters are taken as given for the settlement problem. Write
\begin{align}
q_r^w &= \sum_i c_iq_r^{(i)},&
n_r &= \sum_i q_r^{(i)},\\
L_\alpha(r) &= \sum_{e\in P_r}\alpha_e l_e,&
B(r) &= \sum_{e\in P_r}\beta_e .
\end{align}
The route's mileage revenue and fixed-surcharge revenue are, respectively,
\begin{equation}\label{eq:tariff}
t_r^m=q_r^wL_\alpha(r),\qquad t_r^f=n_r B(r).
\end{equation}
Total revenue is $t=\sum_r(t_r^m+t_r^f)$. Flow profiles may represent observed route records, assignments supplied by an external traffic model, or scenarios. For a tree, the route of each origin--destination pair is determined by topology. Origin--destination flows by class then suffice. On a general network, an origin--destination matrix alone may leave the segment use undetermined.

The allocation formulas and cooperative-game statements below also apply to exogenous route revenues specified directly. For this formulation we require $t_r^c\ge 0$ and $t_r^c=0$ whenever the route has no segment with positive weight for component $c$. This separates the settlement rule from the tariff formula and allows comparison with the exogenous toll matrix in \citet{wu2024}. Both the traffic-flow domain in \eqref{eq:tariff} and the direct-revenue domain permit combining profiles on a common route family and setting individual routes to zero. The proofs use these operations.

\subsection{Weighted proportional sharing and the equal-sharing benchmark}

For $c\in\{m,f\}$, define the weights and the sets on which they are positive by
\begin{equation}\label{eq:support}
w_e^m=\alpha_e l_e,\qquad w_e^f=\beta_e,\qquad
E_c=\{e\in E:w_e^c>0\},\qquad P_r^c=P_r\cap E_c .
\end{equation}
For any $A\subseteq E$, let $w^c(A)=\sum_{e\in A}w_e^c$. A positive-revenue route for component $c$ has $P_r^c\ne\varnothing$. The WPS allocation is
\begin{equation}\label{eq:wps}
\WPS_e^c=\sum_{\substack{r\in\R:\ t_r^c>0\\e\in P_r^c}}
t_r^c\frac{w_e^c}{w^c(P_r^c)},
\qquad
\WPS_e=\WPS_e^m+\WPS_e^f .
\end{equation}
An empty sum is zero. This convention assigns zero to a segment outside $E_c$ for component $c$ and treats an entirely absent surcharge component. Each nonzero term is a share of one route's revenue, so WPS preserves revenue at both the route and component levels. Under the tariff model, the route totals in \eqref{eq:tariff} cancel the route denominators in \eqref{eq:wps}, giving
\begin{equation}\label{eq:operational}
\WPS_e=\alpha_e l_e\sum_{r:e\in P_r}q_r^w
       +\beta_e\sum_{r:e\in P_r}n_r .
\end{equation}
Thus, when these two measures of segment use are available directly and the tariff is additive as in \eqref{eq:tariff}, they suffice to compute WPS. Route data remain the input for the coalition values and counterfactuals studied below.

The route extension of the equal-sharing benchmark is
\begin{equation}\label{eq:ses}
\SES_e=\sum_{r:e\in P_r}\frac{t_r^m+t_r^f}{|P_r|}.
\end{equation}
WPS agrees with this rule if each positive-revenue route has a common strictly positive weight across all its segments for the relevant component. This condition holds, for example, when weighted lengths are constant and surcharges are either constant and positive or identically zero. If a route generates a surcharge but includes a segment with zero surcharge, equal sharing of the combined payment and WPS can differ even when all positive surcharge weights are equal. When $G$ is a line and its segments are indexed in order, the route from the entrance of segment $h$ to the exit of segment $k$ has segment set $\{h,\dots,k\}$. Matching its total revenue to $t_{hk}$ in \citet{wu2024} gives their SES formula term by term.

\subsection{Settlement inputs and aggregation}

There are two levels of data in the model. A tariff-derived route record lists the route, vehicle volumes by class and the attributes of its segments. It gives both components through \eqref{eq:tariff}. A revenue record can instead give the realised amounts for each route and component directly. In either case the network may admit unused routes. A route in the admissible family with zero recorded revenue contributes zero and remains available as an input to a different profile.

A single OD entry can be used as a route entry only when the relevant route is known. When several routes are used for one OD pair, assigning the whole OD total to one route changes the segment-use information, even if the aggregate payment is preserved. By contrast, multiple records that have the same segment set can be combined by adding their component revenues. Their contributions to WPS and to coalition worths depend on the same weight ratios and eligibility sets. This aggregation can be carried out separately for each accounting period before combining the periods.

Equation~\eqref{eq:operational} also shows the data required for an implementation under additive tariffs. For mileage, it is enough to know the sum of class-weighted flows on each segment. For surcharges, it is enough to know the total count on each segment. A single pass through the records can accumulate these two quantities. Route-level records are still needed when one wants to evaluate the coalition games, compare WPS with route SES, or determine which other segment allocations change when a route is cancelled. The numerical illustration distinguishes these uses of the data.

\subsection{Scope of the settlement model}\label{sec:settlement-scope}

The model applies to a finite connected highway network, including networks with cycles, over a specified settlement period. It takes as input the simple routes used during that period and either their class-specific flows under the tariff in \eqref{eq:tariff} or their recorded nonnegative revenues by component. Direction-specific records can be included as separate routes. The communication graphs in this paper are undirected. The segment weights and the eligibility of each route for settlement are fixed within a profile and are common to the compared profiles in the axioms that require fixed attributes.

Coalition worth is evaluated using the specified routes. In the mileage game a coalition earns the revenues of the specified routes it contains. The surcharge game evaluates claims to already generated revenue using the eligible part of each route. This interpretation allows a positive-surcharge segment to receive a settlement even if travel on the full route also required a zero-surcharge segment. For the closure comparisons, routes using the closed segment are cancelled and the other route records retain their original revenues. The resulting statements concern a settlement at given traffic and tariff inputs.

\section{Cooperative-game representation}\label{sec:game}

Segments are the players. For each component $c$ we use the game $(E_c,v^c)$ given by
\begin{equation}\label{eq:game}
v^c(S)=\sum_{\substack{r\in\R:\ t_r^c>0\\P_r^c\subseteq S}}t_r^c,\qquad S\subseteq E_c .
\end{equation}
In the mileage game a coalition earns the revenue of each specified route whose full set of segments it contains. For surcharges, we first identify each route's fixed revenue from the full route record and then project the set of players entitled to share it onto $E_f$. In particular, $v^f$ represents surcharge claims for settlement. A set of surcharge-eligible segments may be separated by segments with $\beta_e=0$, so the projected game need not be a graph-restricted game on the induced road adjacency graph.

Using the unanimity-game expansion in \eqref{eq:dividend-expansion}, we can identify the dividends of these games directly from the route records.

\Needspace{9\baselineskip}
\begin{proposition}[Route dividends]\label{prop:dividends}
For each $c\in\{m,f\}$ and nonempty $A\subseteq E_c$,
\begin{equation}\label{eq:dividends}
\Delta_{v^c}(A)=\sum_{\substack{r:\ t_r^c>0\\P_r^c=A}}t_r^c .
\end{equation}
For the full-route game on $E$, with $P_r$ in place of $P_r^c$, the same statement holds with the full segment sets. In particular, a coalition that is disconnected or is not the segment set of a simple route has zero dividend in the full-route game. Conversely, when route revenues are specified directly, any TU-game with nonnegative dividends supported on the segment sets of admissible routes can be realized as a full-route game.
\end{proposition}

Different routes can have the same eligible set, especially after projection. Their revenues then enter the same dividend. Formula~\eqref{eq:dividends} is valid even when the union of used routes contains cycles. It follows from the route-based definition of coalition worth.

\begin{theorem}[Weighted Shapley representation]\label{thm:representation}
For each $e\in E$, $\WPS_e$ is the sum of $\Sh^{\,w^m|_{E_m}}_e(v^m)$ and $\Sh^{\,w^f|_{E_f}}_e(v^f)$, where a component is extended by zero to segments outside its player set. An empty player set contributes the zero allocation.
\end{theorem}

\begin{corollary}[Route SES and the Shapley value]\label{cor:ses}
The route SES allocation in \eqref{eq:ses} is the ordinary Shapley value of the full-route game on $E$ with route revenues $t_r^m+t_r^f$. On a line this specializes to the segments allocation game, interval dividends and SES representation of \citet{wu2024}.
\end{corollary}

\subsection{Physical connectivity and surcharge eligibility}

The full-route game is component additive on $L(G)$, since a connected route contained in a coalition lies in a single connected component. Proposition~\ref{prop:components} and the route-dividend formula agree on the implication for disconnected coalitions. The route formula gives a stronger restriction: even a connected coalition has zero dividend unless a recorded route has precisely that segment set. As a result, a connected coalition containing a physical cycle can have positive worth and zero dividend.

For a route consisting of three consecutive segments with surcharges $(2,0,1)$, suppose that the route generates a fixed-surcharge revenue of 90. The full-route surcharge game assigns this revenue to the unanimity game on all three segments. The projected surcharge game has dividend 90 on the pair of outer segments and WPS assigns them 60 and 30. The pair is disconnected in the induced graph on positive-surcharge segments. If that projected game were graph restricted on the induced graph, the worth of the pair would be the sum of two zero singleton worths. The projected game is therefore interpreted as a settlement game for specified route claims.

The distinction between physical routes and communication links also explains the choice of perturbation in Section~\ref{sec:axioms}. As a simple graph-game example, suppose that three segment players form a triangle in the communication graph and that the game is $u_{\{1,2\}}$. After deletion of the communication link between 1 and 2, the grand coalition can still connect them through 3, so the resulting graph restriction has worth one for the grand coalition. Under route cancellation, deleting the only specified contribution that uses both 1 and 2 makes all coalition worths zero. The second characterization below is formulated in terms of route cancellation.

\section{Two axiomatic characterizations}\label{sec:axioms}

Fix a network, a route family and the segment attributes. An allocation rule assigns a nonnegative settlement vector to each admissible revenue profile. We state the axioms for pure components and use separability to address mixed profiles. A pure profile has revenue in only one of the two components. For a mixed profile $P$, write $P^m$ and $P^f$ for its pure mileage and pure surcharge counterparts. Each counterpart keeps the corresponding route revenues of $P$. In the tariff model, these are obtained by suppressing the other charge while preserving the route flows. For a pure component we denote the rule by $\psi^c$ and the route revenues by $t_r^c(P)$.

\paragraph{Separability.}
For every profile $P$, the rule satisfies $\psi(P)=\psi^m(P^m)+\psi^f(P^f)$.

\paragraph{Zero surcharge weight.}
In a pure surcharge problem, $\psi_e^f(P)=0$ whenever $\beta_e=0$.

\subsection{Additivity and proportional treatment}

\paragraph{Route additivity.}
For two pure profiles $P,P'$ with the same component and fixed attributes, let $P+P'$ add the corresponding route revenues. Then $\psi^c(P+P')=\psi^c(P)+\psi^c(P')$. With the tariff schedule held fixed, this operation corresponds to adding route flows.

\paragraph{Efficiency.}
In every pure problem, $\sum_{e\in E}\psi_e^c(P)=\sum_r t_r^c(P)$.

\paragraph{Inessential segment property.}
If a segment belongs to no positive-revenue route in a pure problem, it receives zero in that component.

\paragraph{Proportional treatment of commonly used segments.}
In a pure problem with positive total revenue, if $i,j\in E_c$ both belong to every positive-revenue route, then
\[
\frac{\psi_i^c(P)}{w_i^c}=\frac{\psi_j^c(P)}{w_j^c}.
\]
The axiom compares segments whose presence is common to all of the route revenue under consideration. WPS also satisfies this equality for any two eligible segments that have identical membership across the positive-revenue routes.

\begin{theorem}[Characterization by route additivity]\label{thm:additivity}
WPS is the unique allocation rule satisfying separability, zero surcharge weight, route additivity, efficiency, the inessential segment property and proportional treatment of commonly used segments.
\end{theorem}

On a line with one unweighted component, these axioms reduce to the additivity, efficiency, inessential segment and segment symmetry requirements of \citet{wu2024}. With equal weights, common use calls for equal allocations. With heterogeneous weights, it calls for equal allocations per unit of the specified charging basis.

\subsection{Sub-highway efficiency and average balanced contributions}

Fix a pure profile and a component $c$. A sub-highway is a nonempty set $C\subseteq E$ that is connected in the segment adjacency graph and closed under positive-revenue routes. Thus if $t_r^c(P)>0$ and $P_r\cap C\ne\varnothing$, then $P_r\subseteq C$. Two road segments are adjacent when they share an endpoint. This definition uses the full physical route $P_r$ for both components. A zero-surcharge segment may therefore connect two eligible segments within the same sub-highway.

Let $P^{-j}$ be the profile obtained by setting to zero the revenue of every specified route containing segment $j$ and retaining the revenue of every other route. In the flow model this means setting the corresponding class flows to zero. The underlying network, route family and segment attributes are the same in $P$ and $P^{-j}$. The operation represents cancellation of trips on the affected routes. It can be evaluated using the supplied profile, without assigning those trips to alternative routes.

\paragraph{Sub-highway efficiency.}
For every sub-highway $C$ of the pure profile $P$,
\begin{equation}\label{eq:subeff}
\sum_{e\in C}\psi_e^c(P)=\sum_{r:P_r\subseteq C}t_r^c(P).
\end{equation}

\paragraph{Weighted average balanced contributions.}
For a sub-highway $C$, write $H=C\cap E_c$ for its eligible segments. For every $i\in H$,
\begin{equation}\label{eq:abc}
\sum_{j\in H\setminus\{i\}}
\frac{\psi_i^c(P)-\psi_i^c(P^{-j})}{w_i^c}
=
\sum_{j\in H\setminus\{i\}}
\frac{\psi_j^c(P)-\psi_j^c(P^{-i})}{w_j^c}.
\end{equation}
For $|H|\ge 2$, dividing both sides by $|H|-1$ gives the average comparison. It equates the average normalized effect on $i$ of cancelling routes through each other eligible segment with the average normalized effect on those segments of cancelling routes through $i$.

\begin{theorem}[Characterization by average balanced contributions]\label{thm:abc}
WPS is the unique allocation rule satisfying separability, zero surcharge weight, sub-highway efficiency and weighted average balanced contributions.
\end{theorem}

The sub-highway efficiency condition adapts component efficiency in communication games \citep{myerson1977}. The average comparison uses the idea of reciprocal allocation effects underlying balanced contributions and its average formulation \citep{myerson1980,li2026}. Here the comparison concerns changes in the specified route profile. In particular, the proof uses connected groups of full physical routes and makes comparisons only among their positive-weight segments. The fixed-surcharge characterization therefore also covers routes on which zero-surcharge segments separate positive-surcharge segments.

\section{Proportional and compensated sharing on general networks}\label{sec:sps-scs}

The route model also allows us to extend the two rules of \citet{soto2026}. For SPS we give a cooperative-game representation and an axiomatic characterization. For SCS we specify and characterize a rule based on rooted communication trees and examine its relation to the linear-highway formula.

\subsection{Segments proportional sharing and the \texorpdfstring{$\tau$}{tau} value}

We first work with a single stream of revenue. Let $N$ be its set of eligible segments, $t_r\ge 0$ the revenue of route $r$, and $A_r\subseteq N$ its eligible segment set. Use $(N,A_r,t_r)=(E_c,P_r^c,t_r^c)$ for component $c\in\{m,f\}$. For a pooled comparison with the original SPS rule, use $(E,P_r,t_r^m+t_r^f)$. In either case $A_r$ is nonempty whenever $t_r>0$. Define
\begin{align}
s_e &= \sum_{r:A_r=\{e\}}t_r, &
d_e &= \sum_{\substack{r:e\in A_r\\ |A_r|\ge 2}}t_r, &
D &= \sum_{r:|A_r|\ge 2}t_r . \label{eq:sps-data}
\end{align}
The amount $s_e$ is directly attributable to $e$ in the settlement game. A physical route with several segments may contribute to $s_e$ in the surcharge game if only $e$ has a positive surcharge. The amount $D$ is the joint revenue available for distribution. The quantities $d_e$ record each eligible segment's participation in that joint revenue. A route with several eligible segments is counted once in $D$ and once for each of its eligible segments in $\sum_e d_e$.

The route version of Segments Proportional Sharing is
\begin{equation}\label{eq:sps-rule}
\operatorname{SPS}_e=s_e+\frac{d_e}{\sum_{j\in N}d_j}\,D,\qquad e\in N,
\end{equation}
with the second term set to zero if $D=0$. We extend the vector by zero outside $N$. The componentwise SPS rule sums the two vectors obtained from mileage and surcharge revenues. The weights of the charging system enter via the revenues $t_r^c$ and the eligible sets $P_r^c$. Once those quantities are fixed, the allocation of joint revenue is proportional to $d_e$.

For a TU game $v$ on $N$, the utopia payoff is $M_e=v(N)-v(N\setminus\{e\})$. The minimum right is
\[
m_e=\max_{\substack{S\subseteq N\\ e\in S}}
\left\{v(S)-\sum_{j\in S\setminus\{e\}}M_j\right\}.
\]
The $\tau$ value, when defined, is the efficient point on the segment between the vectors $m$ and $M$ \citep{tijs1986}. In our games the minimum rights and utopia payoffs have direct revenue interpretations.

\begin{theorem}[Route SPS and the \texorpdfstring{$\tau$}{tau} value]\label{thm:sps-tau}
For the game $v=\sum_{r:t_r>0} t_r u_{A_r}$ on $N$, route SPS equals $\tau(v)$. Componentwise SPS therefore equals $\tau(v^m)+\tau(v^f)$, with zero extension outside each component's eligible set. For a linear highway with pooled revenue and full route sets, \eqref{eq:sps-rule} recovers the original SPS formula and its $\tau$-value representation.
\end{theorem}

The componentwise rule and the pooled rule can differ because the $\tau$ value need not be additive. For example, consider two unit revenues associated with disjoint supports of sizes two and three. The pooled rule gives each of the five segments $2/5$. If the two revenues belong to different components and are allocated separately, each support receives its entire unit. The example also shows that a coalition can receive less than its worth under pooled SPS. In this case the two-segment coalition receives $4/5$ even though it generates one unit on its own.

\subsection{An axiomatic characterization of route SPS}\label{sec:sps-axioms}

Use the separability, zero surcharge weight, efficiency and inessential segment properties from Section~\ref{sec:axioms}. For each pure component we add two requirements adapted from \citet{soto2026}. For the pooled rule we apply the requirements on $N=E$ and omit separability and zero surcharge weight.

\paragraph{Weak toll covariance.}
Suppose that, between two pure profiles with the same network, attributes and route family, all revenues with $|A_r|\ge 2$ are multiplied by a common factor $b\ge 0$ and the singleton revenues satisfy $s'_e=b s_e+a_e$ for every $e\in N$. Whenever both profiles are admissible, the allocation satisfies $\psi'_e=b\psi_e+a_e$ on $N$ and $\psi'_e=b\psi_e$ outside $N$. This requires a change in directly attributable revenue to accrue to the corresponding segment.

\paragraph{Proportionality for joint revenue.}
If all singleton revenues are zero, then for every pair $e,j\in N$ with $d_e,d_j>0$, the allocations satisfy $d_j\psi_e=d_e\psi_j$. Thus the allocation per unit of joint-revenue participation is equal among the eligible segments that participate. The cross-multiplied equation remains meaningful for a candidate rule with zero payoffs.

\begin{theorem}[Characterization of route SPS]\label{thm:sps-axioms}
Componentwise SPS is the unique allocation rule satisfying separability, zero surcharge weight, efficiency, the inessential segment property, weak toll covariance and proportionality for joint revenue. The pooled route SPS rule is uniquely characterized by the last four properties in the single-stream model with full route sets.
\end{theorem}

The characterization uses covariance instead of route additivity. Once directly attributable revenues have been removed, efficiency and proportionality determine a common division of the joint revenue. The scale of the joint pool depends on the whole profile, so the rule does not require that each route distribute its revenue separately.

\subsection{Segments compensated sharing using rooted communication trees}\label{sec:scs}

For the mileage component, use the line graph $L(G)$ as the communication graph $H_m$. Every full route $P_r$ is connected in it. For the surcharge component, define $H_f$ on $E_f$. Join two distinct eligible segments when a path in $L(G)$ connects them with all internal vertices outside $E_f$. In particular, adjacent eligible segments are joined. This convention preserves communication through zero-surcharge segments without making them eligible to receive surcharge revenue. Consecutive eligible segments along a physical route are adjacent in $H_f$, so $P_r^f$ is connected. If $E_f$ is empty, the surcharge allocation is zero.

For a nonempty component player set, let $\mathcal B(H_c)$ be the admissible rooted spanning trees of $H_c$ in the sense of \citet{herings2010}. These are the rooted spanning trees obtainable by depth-first search \citep{baron2011}. The construction extends the cycle-free average tree solution of \citet{herings2008}. They can be described recursively. After choosing a root, remove it and consider the resulting connected components. In each component choose a new root adjacent to the removed root and repeat within that component. Let $\rho(B)$ be the root of $B$, and let $D_e(B)$ be the set consisting of $e$ and its descendants. The hierarchical allocation in tree $B$ is
\begin{equation}\label{eq:scs-hierarchy}
h^B_e(v^c)=v^c\bigl(D_e(B)\bigr)
-\sum_{j:\,j\text{ is a child of }e\text{ in }B}v^c\bigl(D_j(B)\bigr).
\end{equation}
A weighted extension of Segments Compensated Sharing is
\begin{equation}\label{eq:scs-rule}
\operatorname{SCS}_e^c =
\frac{\sum_{B\in\mathcal B(H_c)} w_{\rho(B)}^c\, h_e^B(v^c)}
{\sum_{B\in\mathcal B(H_c)} w_{\rho(B)}^c},\qquad e\in E_c.
\end{equation}
The total is $\operatorname{SCS}_e^m+\operatorname{SCS}_e^f$, with zero extension outside each eligible set. The formula assigns weight to each tree according to its root. If a segment is the root of several admissible trees, each of those trees receives its weight in the average. Constant weights yield the ordinary average tree solution for the specified graph.

The rule can be computed route by route. In each admissible tree, start at its root and, if the root is outside the eligible route set, descend into the unique child component containing that set. Stop at the first member of the set. The corresponding segment receives the route revenue in that tree. The average therefore shares each route's revenue among its eligible segments. In the pooled model on a line, uniform weights and the path communication graph of consecutive road segments recover the SCS formula of \citet{soto2026}. Appendix~\ref{app:scs} verifies this specialization.

\subsection{A recursive characterization of compensated sharing}\label{sec:scs-axioms}

We give a characterization on an enlarged settlement domain that includes problems with a designated root segment. This allows the axioms to describe how allocations are passed to smaller communication components. The domain consists of connected induced subgraphs of the given communication graph, together with nonnegative revenues on connected route supports. If $K$ is a connected vertex subset, the subproblem on $H[K]$ retains precisely the revenues with eligible supports contained in $K$. The communication graph is restricted by induction from the given graph. In particular, surcharge shortcuts are specified before forming the subproblems. These are settlement subproblems and their construction does not assign vehicles to other routes.

For a connected graph $H$ and $i\in V(H)$, let $n_i(H)$ denote the number of its admissible spanning trees rooted at $i$. This number is determined by the communication graph and is independent of toll revenues. For a component $K$ of $H-i$, let $J_i(K)$ be the vertices of $K$ adjacent to $i$ in $H$. The admissible roots for the branch $K$ are exactly the members of $J_i(K)$. Define the probabilities
\begin{equation}\label{eq:scs-lotteries}
p_i(H,w)=\frac{w_i n_i(H)}{\sum_{z\in V(H)}w_z n_z(H)},\qquad
p_{ij}(H,K)=\frac{n_j(H[K])}{\sum_{z\in J_i(K)}n_z(H[K])},
\quad j\in J_i(K).
\end{equation}
The first probability selects the root of the whole settlement and weights it by the number of admissible trees available at that root. The second selects a root for a branch and weights each choice by its number of admissible continuations. Thus the construction weights the initial root by the segment attribute and treats tree completions equally conditional on that root. The branch probabilities do not depend on the allocation rule.

Let $x^i(H,v)$ be the vector conditional on the designation of root $i$, and let $x(H,v;w)$ be the allocation before a root is designated. We impose the following requirements on this pair of rules.

\paragraph{Rooted efficiency.}
For every designated root $i$, the conditional allocation distributes the entire available revenue, $\sum_{e\in V(H)}x^i_e(H,v)=v(V(H))$.

\paragraph{Branch settlement consistency.}
For every component $K$ of $H-i$, the restriction of $x^i(H,v)$ to $K$ equals the expected allocation obtained by settling the subproblem on $K$ with a root chosen from $J_i(K)$:
\begin{equation}\label{eq:branch-consistency}
x^i(H,v)|_K=\sum_{j\in J_i(K)}p_{ij}(H,K)\,x^j(H[K],v|_K).
\end{equation}
Here $v|_K$ is the restriction of $v$ to subsets of $K$. The axiom compares conditional settlements across graphs of different sizes. It leaves the designated root's allocation to be determined by efficiency.

\paragraph{Consistency with the initial root lottery.}
The allocation without a designated root is the expectation of the conditional allocations under the initial probabilities in \eqref{eq:scs-lotteries}:
\[
x(H,v;w)=\sum_{i\in V(H)}p_i(H,w)\,x^i(H,v).
\]

\begin{theorem}[Recursive characterization of weighted compensated sharing]\label{thm:scs-recursive}
On the domain above, rooted efficiency, branch settlement consistency and consistency with the initial root lottery uniquely determine the conditional and unconditional rules. The conditional rule averages the hierarchical allocations of the admissible trees rooted at the designated segment. The unconditional rule is the weighted average in \eqref{eq:scs-rule}. With constant weights it is the average tree solution of \citet{herings2010}.
\end{theorem}

The characterization is conditional on the communication graph and the probabilities in \eqref{eq:scs-lotteries}. It determines allocations through settlement on smaller components. In comparison, the hierarchy axiom of \citet{baron2011} uses tree frequencies to prescribe relative allocations on unanimity games. Our recursive axioms give a characterization on the route-settlement domain, including general graphs, while taking the choice of tree probabilities as part of the settlement protocol.

\section{Numerical illustration on a cyclic highway network}\label{sec:example}

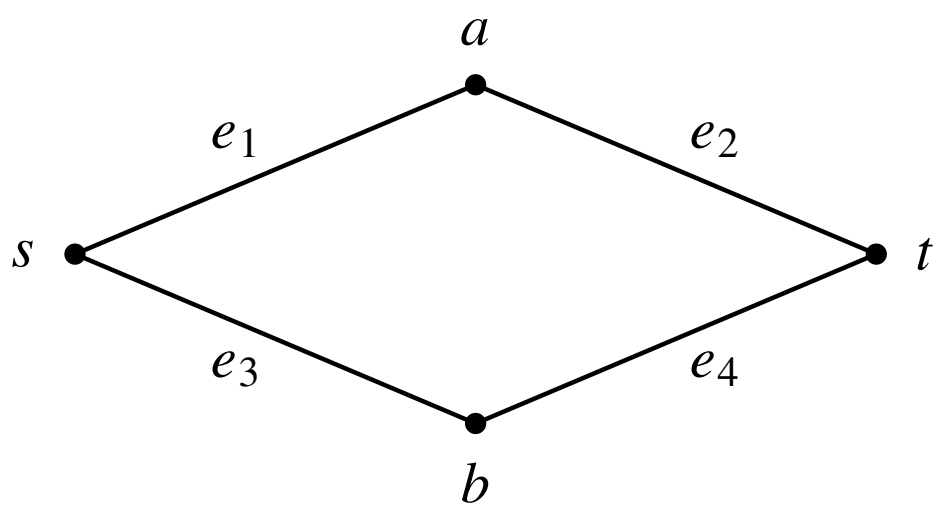
\begin{figure}[H]
\centering
\begin{tikzpicture}[line width=.85pt,every node/.style={font=\small}]
\coordinate (s) at (0,0);
\coordinate (a) at (2.65,1.12);
\coordinate (t) at (5.3,0);
\coordinate (b) at (2.65,-1.12);
\draw (s) -- node[above left, inner sep=2pt] {$e_1$} (a)
          -- node[above right, inner sep=2pt] {$e_2$} (t)
          -- node[below right, inner sep=2pt] {$e_4$} (b)
          -- node[below left, inner sep=2pt] {$e_3$} (s);
\fill (s) circle (2pt) node[left=3pt] {$s$};
\fill (a) circle (2pt) node[above=3pt] {$a$};
\fill (t) circle (2pt) node[right=3pt] {$t$};
\fill (b) circle (2pt) node[below=3pt] {$b$};
\end{tikzpicture}
\caption{Two routes between $s$ and $t$ on a network with a cycle.}
\label{fig:cycle}
\end{figure}

Consider the four-segment network in Figure~\ref{fig:cycle}. Two routes connect $s$ and $t$. A third route connects $a$ and $b$ via $s$. The illustration uses scenario inputs. The mileage rates of cars and trucks are 1 and 2, respectively. The segment weights and surcharges are reported in Table~\ref{tab:edges}.

\begin{table}[H]
\centering
\caption{Segment attributes in the illustrative tariff.}
\label{tab:edges}
\begin{tabular}{lrr}
\toprule
Segment & Weighted length $\alpha_e l_e$ & Fixed surcharge $\beta_e$\\
\midrule
$e_1=sa$ & 2 & 0\\
$e_2=at$ & 4 & 2\\
$e_3=sb$ & 3 & 1\\
$e_4=bt$ & 3 & 1\\
\bottomrule
\end{tabular}
\end{table}

In scenario A, 60 cars and 20 trucks use $s$--$a$--$t$, while 40 cars and 20 trucks use $s$--$b$--$t$. Scenario B shifts 20 cars and 10 trucks from the first route to the second. The $a$--$s$--$b$ flow is held fixed. Each $s$--$t$ route has weighted length 6 and total surcharge 2. The shift therefore preserves the revenue paid per vehicle of each class. Both scenarios have the same origin--destination volumes by class, the same aggregate mileage and surcharge revenues, and the same total revenue of 1,590. Table~\ref{tab:flows} gives the corresponding route totals.

\begin{table}[H]
\centering
\caption{Scenario flows and total toll revenue by route.}
\label{tab:flows}
\begin{tabular}{lrrrrrr}
\toprule
& \multicolumn{3}{c}{Scenario A} & \multicolumn{3}{c}{Scenario B}\\
\cmidrule(lr){2-4}\cmidrule(lr){5-7}
Route & Cars & Trucks & Revenue & Cars & Trucks & Revenue\\
\midrule
$s$--$a$--$t$ & 60 & 20 & 760 & 40 & 10 & 460\\
$s$--$b$--$t$ & 40 & 20 & 600 & 60 & 30 & 900\\
$a$--$s$--$b$ & 20 & 10 & 230 & 20 & 10 & 230\\
\midrule
Total & 120 & 50 & 1,590 & 120 & 50 & 1,590\\
\bottomrule
\end{tabular}
\end{table}

\begin{table}[H]
\centering
\caption{Settlement by segment under WPS and route SES.}
\label{tab:results}
\begin{tabular}{lrrrr}
\toprule
& \multicolumn{2}{c}{Scenario A} & \multicolumn{2}{c}{Scenario B}\\
\cmidrule(lr){2-3}\cmidrule(lr){4-5}
Segment & WPS & SES & WPS & SES\\
\midrule
$e_1$ & 280 & 495 & 200 & 345\\
$e_2$ & 560 & 380 & 340 & 230\\
$e_3$ & 450 & 415 & 600 & 565\\
$e_4$ & 300 & 300 & 450 & 450\\
\midrule
Total & 1,590 & 1,590 & 1,590 & 1,590\\
\bottomrule
\end{tabular}
\end{table}

Table~\ref{tab:results} shows the effects of both the rule and the route split. In scenario A, the $s$--$a$--$t$ route generates 600 in mileage revenue and 160 in surcharge revenue. WPS divides the mileage revenue in the ratio $2:4$ and assigns the whole surcharge to $e_2$. SES allocates 380 to each of the two segments. The traffic shift in scenario B moves revenue from $e_1$ and $e_2$ to $e_3$ and $e_4$ under both rules. The allocations differ across scenarios despite the identical OD totals, including the OD toll revenue totals. In this example, the difference is attributable to the changed segment use within one OD pair.

\subsection{Dividends and route cancellation in the example}

For scenario A, the positive route contributions to each component are shown in Table~\ref{tab:route-components}. Mileage dividends occur on the three full route sets. Surcharge dividends occur on the three eligible sets. For example, $e_1$ is needed to complete route $s$--$a$--$t$, but its surcharge weight is zero. The full physical route generates 160 of surcharge revenue, which appears as a singleton dividend on $\{e_2\}$ in the projected game.

\begin{table}[H]
\centering
\caption{Scenario A dividends by physical route and surcharge eligibility.}
\label{tab:route-components}
\begin{tabular}{lrrl}
\toprule
Physical segment set & Mileage & Surcharge & Surcharge support\\
\midrule
$\{e_1,e_2\}$ & 600 & 160 & $\{e_2\}$\\
$\{e_3,e_4\}$ & 480 & 120 & $\{e_3,e_4\}$\\
$\{e_1,e_3\}$ & 200 & 30 & $\{e_3\}$\\
\bottomrule
\end{tabular}
\end{table}

In scenario A, componentwise SPS assigns $(400,300,340,240)$ in mileage revenue and $(0,160,90,60)$ in surcharge revenue. Its total allocation is $(400,460,430,300)$. In the pooled game every positive route uses exactly two segments, so pooled SPS allocates half of each route's total revenue to each member and coincides with route SES in Table~\ref{tab:results}. The componentwise comparison makes the treatment of the singleton surcharge supports visible.

The grand coalition has mileage worth 1,280, the sum of the three mileage entries. Its mileage dividend is zero because no individual route uses all four segments. The coalition $\{e_1,e_2,e_3\}$ has mileage worth 800 and also has zero mileage dividend, since the revenues it contains are already assigned to $\{e_1,e_2\}$ and $\{e_1,e_3\}$. The same reasoning applies even when the edges of a connected coalition form a simple path that is absent from the route records.

The mileage comparison between $e_1$ and $e_2$ illustrates the cancellation axiom. Cancelling routes through $e_2$ removes only the first route, reducing the mileage allocation of $e_1$ by 200. Cancelling routes through $e_1$ also removes the third route, but the latter generates no revenue for $e_2$. The mileage allocation of $e_2$ thus falls by 400. The normalized losses are the same, since $200/2=400/4=100$. The comparison depends on the route shared by the two segments. Cancelled routes used by just one of the two do not enter this pair of cross-effects.

The surcharge comparison between $e_3$ and $e_4$ illustrates the same comparison for a different revenue component. Each loses 60 of surcharge allocation when routes through the other segment are cancelled, and both have surcharge weight 1. Segment $e_1$ remains a physical member of the third route, though it is excluded from comparisons of surcharge allocations per unit weight.

\subsection{What the illustration identifies}

The two scenarios hold both OD demand and aggregate toll receipts fixed. Their different settlements identify the effect of the route split at the stated tariff. They also show why testing a settlement rule against aggregate receipts alone is insufficient. For example, a rule that credited the $s$--$t$ revenue to one predetermined route would give the same allocation under both scenarios and would miss the change in segment use. The illustration is a constructed comparison of allocation mechanisms. An application to a particular highway would require route records or a separately justified estimate of the route split.

\section{Conclusions and further extensions}\label{sec:conclusion}

We extended the SES approach of \citet{wu2024} from a line to a general highway network with given route flows. Route SES retains equal division of each route's toll among the segments it uses. WPS extends that principle by distinguishing mileage weights from surcharge weights. It has a weighted Shapley representation for each component and two axiomatic characterizations. On a line with equal weights, the model recovers the 2024 SES rule, its route-interval dividend formula, its Shapley representation and its characterization based on additivity, efficiency, inessential segments and segment symmetry. Cycles enter through the coexistence of multiple routes. The settlement rule and the proofs apply after their flows have been specified.

\paragraph{Extensions of the other 2024 methods.}
The Exits Equal Sharing and Entrances Equal Sharing methods of \citet{wu2024} can be extended by the same allocation framework once the settlement agreement specifies the eligible segments for each exit or entrance. For example, let $U(x)$ be a specified set of upstream segments eligible for revenue collected at exit $x$. Aggregate the payments collected at $x$ and divide that amount among $U(x)$, equally or in proportion to specified positive weights. An analogous downstream set $D(a)$ allocates the payments collected at entrance $a$. On a directed line, choosing the prefix ending at the exit segment or the suffix beginning at the entrance segment reproduces Exits Equal Sharing or Entrances Equal Sharing, respectively, under equal weights. A general network can support different eligibility policies, especially where routes form cycles. Establishing counterparts of the original axioms for particular eligibility policies remains an open question.

\paragraph{The 2026 results.}
We established the SPS--$\tau$ equivalence and a route-based axiomatic characterization of SPS on general networks. The pooled model recovers the SPS result of \citet{soto2026} on a line, and the componentwise model respects surcharge eligibility. For SCS, the weighted average of DFS tree allocations recovers the linear rule in the pooled equal-weight case. Its recursive characterization uses efficiency conditional on a designated root, consistency across the resulting branch settlements, and averaging under a specified initial root lottery.

\paragraph{Scope and open questions.}
On networks with several possible routes for the same OD pair, OD totals alone may leave the segment allocation undetermined. Applying these rules in that situation requires route records, route shares supplied by an external traffic assignment, or explicit scenarios. When tolls, congestion or closures lead drivers to change routes, a model of reassignment is needed to obtain the new profile. Coalition worth under alternative route choices is a different modelling problem. Extending the rules to games in which coalitions can reroute traffic, or to settlements with negative route revenues, requires separate analysis.

For SCS, the recursive characterization treats the communication graph and root lotteries as inputs. A further question is whether observable changes in toll revenues or route closures can identify those inputs. The component fairness axiom of \citet{soto2026} compares the average losses on two sides of a severed connection. A single closure in a general graph may leave the communication graph connected. Even a physical three-arm tree has a triangular segment communication graph, so deleting one communication link creates no partition. Extending that fairness principle would require a different comparison and an economic justification of the surcharge shortcuts and tree probabilities.

\clearpage
\appendix
\section{Proofs}\label{app:proofs}

\subsection{Dividend identities and component additivity}\label{app:components}

To see that \eqref{eq:mobius} inverts \eqref{eq:dividend-expansion}, substitute the first formula into the second and exchange the finite sums. For each $D\subseteq A$, the coefficient of $\Delta_v(D)$ becomes
\[
\sum_{B:\,D\subseteq B\subseteq A}(-1)^{|A|-|B|}
=(1-1)^{|A\setminus D|}.
\]
This coefficient is one if $D=A$ and zero otherwise. Equivalently, the equations $v(S)=\sum_{A\subseteq S}\Delta_v(A)$ determine the coefficients recursively by coalition size.

For Proposition~\ref{prop:components}, first suppose that the nonzero dividends are supported on connected coalitions. Any such coalition contained in $S$ lies in exactly one component of the subgraph induced by $S$. Summing the expansion \eqref{eq:dividend-expansion} separately in those components gives $v(S)=\sum_{C\in\mathcal C_H(S)}v(C)$.

Conversely, suppose $v$ is component additive. Proceed by induction on the size of a disconnected coalition $A$. Every disconnected proper subcoalition has zero dividend by induction. Every connected subcoalition lies in one of the components of $A$. These components are proper subsets of $A$, so the sum of all proper subcoalition dividends equals $\sum_{C\in\mathcal C_H(A)}v(C)=v(A)$. By \eqref{eq:dividend-expansion}, $\Delta_v(A)=0$. The same argument applies to a disconnected pair, which starts the induction. \hfill$\square$

\subsection{Proof of Proposition~\ref{prop:dividends}}

For $S\subseteq E_c$, substituting $u_{P_r^c}(S)=\ind\{P_r^c\subseteq S\}$ into \eqref{eq:game} and collecting terms with the same support gives
\[
v^c=\sum_{r:\,t_r^c>0}t_r^c u_{P_r^c}
=\sum_{\varnothing\ne A\subseteq E_c}
\left(\sum_{\substack{r:\,t_r^c>0\\P_r^c=A}}t_r^c\right)u_A .
\]
Positive-revenue routes have nonempty support. The uniqueness of the unanimity-game expansion therefore gives \eqref{eq:dividends}. The same calculation on $E$ with $P_r$ gives the full-route result. Each $P_r$ is the edge set of a simple path, hence connected in the segment adjacency graph. Coalitions that cannot occur as such a support have zero coefficient. Conversely, for each positive dividend, choose one admissible route with that segment set and assign it revenue equal to the dividend. All other routes have zero revenue. These revenues can all be entered in the mileage component, whose weights are positive on every segment. The resulting full-route game has the prescribed dividends. \hfill$\square$

\subsection{Proof of Theorem~\ref{thm:representation} and Corollary~\ref{cor:ses}}

For $e\in E_c$, the weighted Shapley dividend formula and Proposition~\ref{prop:dividends} give
\begin{align*}
\Sh_e^{\,w^c|_{E_c}}(v^c)
&=\sum_{\substack{A\subseteq E_c\\e\in A}}
  \frac{w_e^c}{w^c(A)}\,\Delta_{v^c}(A)\\
&=\sum_{\substack{r:\,t_r^c>0\\e\in P_r^c}}
  \frac{w_e^c}{w^c(P_r^c)}\,t_r^c
=\WPS_e^c.
\end{align*}
The extension by zero outside $E_c$ agrees with \eqref{eq:wps}. Adding the two components gives the theorem. For the full-route game, use the same dividend calculation with total route revenues and a constant positive weight on $E$. Each route's dividend is then divided by $|P_r|$. With line segments indexed in order, each interval is a unique route, so its dividend is $t_{hk}$ and the resulting Shapley allocation is the linear SES formula. \hfill$\square$

\subsection{Proof of Theorem~\ref{thm:additivity}}

Existence follows from \eqref{eq:wps}. The formula is separable by component and linear in each route revenue at fixed weights. Each positive route distributes its entire revenue, and an inessential or zero-weight segment receives no contribution from that component. For any two eligible segments used on all positive-revenue routes, the terms of their allocations after division by their weights are the same, namely $t_r^c/w^c(P_r^c)$ for each route.

For uniqueness, fix a pure component $c$. A zero-revenue profile yields zero by efficiency and nonnegativity. Any other profile is the sum of profiles that each retain the revenue of just one positive-revenue route. In such a single-route profile for $r$, the inessential segment and zero surcharge weight properties assign zero outside $P_r^c$. All members of $P_r^c$ belong to the sole positive-revenue route. Proportional treatment therefore requires a common constant $\lambda$ with $\psi_e^c=\lambda w_e^c$ for $e\in P_r^c$. Efficiency implies $\lambda=t_r^c/w^c(P_r^c)$. Route additivity now fixes the allocation for every pure profile, and separability fixes it for mixed profiles. \hfill$\square$

\subsection{Proof of Theorem~\ref{thm:abc}}

\paragraph{Existence.}
Separability and zero surcharge weight follow from \eqref{eq:wps}. For any sub-highway $C$, a positive-revenue route is either entirely contained in $C$ or disjoint from it. In the first case its entire component revenue contributes to the sum of allocations in $C$. In the second case its contribution to that sum is zero. This proves sub-highway efficiency.

For distinct $i,j\in E_c$, cancellation of routes through $j$ gives
\begin{equation}\label{eq:pair-loss}
\frac{\WPS_i^c(P)-\WPS_i^c(P^{-j})}{w_i^c}
=\sum_{\substack{r:\,t_r^c(P)>0\\i,j\in P_r}}
   \frac{t_r^c(P)}{w^c(P_r^c)}.
\end{equation}
The right-hand side is symmetric in $i$ and $j$. Summation over the other eligible segments of a sub-highway gives \eqref{eq:abc}.

\paragraph{Uniqueness.}
By separability it is enough to prove the result for a pure component. Fix $c$ and induct on the number of positive-revenue routes. With no such routes, each singleton is a sub-highway and its revenue is zero, so sub-highway efficiency determines the zero allocation.

Now fix a profile $P$ with at least one positive-revenue route and assume uniqueness for every profile with fewer such routes on the same network and route family. Construct a graph with vertex set $E$ by joining segments that are consecutive along at least one positive-revenue route. Each positive-revenue route lies entirely in a single connected component of this graph. Every component $C$ is connected in the physical segment adjacency graph and is closed under positive-revenue routes. It is therefore a sub-highway, and these components partition $E$.

For each such component let $H=C\cap E_c$. All segments of $C\setminus H$ have zero allocation in component $c$. If $H$ has one member, sub-highway efficiency fixes its allocation, and if $H$ is empty, every allocation in $C$ is already fixed. Suppose $|H|\ge 2$. The graph component $C$ has at least two vertices, so every vertex in $C$ is incident to a link of the constructed graph and thus belongs to a positive-revenue route. Cancelling the routes through any $j\in H$ strictly reduces the number of positive-revenue routes. The induction hypothesis gives $\psi^c(P^{-j})=\WPS^c(P^{-j})$.

For $i\in H$, write
\[
d_i=\frac{\psi_i^c(P)-\WPS_i^c(P)}{w_i^c}.
\]
Subtract the average balanced contributions equations for $\psi$ and WPS on $C$. The allocations in all reduced profiles cancel by induction, giving
\[
|H|d_i-\sum_{j\in H}d_j=0
\qquad(i\in H).
\]
Thus $d_i=d$ is constant on $H$. Subtracting the sub-highway efficiency equations yields $d\sum_{i\in H}w_i^c=0$. The sum is positive, so $d=0$. This proves agreement on every component. Induction and separability then yield uniqueness on the full domain. \hfill$\square$

\subsection{Proof of Theorem~\ref{thm:sps-tau}}

In the game $v$, $v(\{e\})=s_e$. Removing $e$ from the grand coalition removes precisely the route revenues whose eligible sets contain $e$. Thus $M_e=s_e+d_e$. For any $S$ containing $e$, each nonempty route support contained in $S$ other than $\{e\}$ contains at least one segment in $S\setminus\{e\}$. Its revenue is counted in at least one of the corresponding utopia payoffs. All route revenues are nonnegative, so
\[
v(S)-\sum_{j\in S\setminus\{e\}}M_j\le s_e .
\]
Equality holds at $S=\{e\}$ and therefore $m_e=s_e$. The gap is $M_e-m_e=d_e$. If $D>0$, then $\sum_e d_e\ge D$ because every joint revenue is counted for all its eligible members. The unique efficient point on the segment between $m$ and $M$ is $s+(D/\sum_e d_e)d$, which gives \eqref{eq:sps-rule}. If $D=0$ then $m=M=s$ and $s$ is efficient. The claim for an empty component follows from zero revenue. On a line, singleton supports correspond to the diagonal toll entries and $d_e$ sums the non-diagonal tolls of all intervals containing $e$. This gives the pooled specialization. \hfill$\square$

\subsection{Proof of Theorem~\ref{thm:sps-axioms}}

Formula~\eqref{eq:sps-rule} is efficient and pays nothing to an eligible segment with neither singleton nor joint revenue. Zero extension and componentwise addition give zero surcharge weight and separability. On a profile with no singleton revenues, the allocation is proportional to $d$. Under the change in the covariance axiom, the vector $d$ and the total $D$ are both multiplied by $b$. For $b>0$ the joint allocation scales by $b$, and for $b=0$ the new joint allocation is zero. The stated transformation of $s$ then gives weak toll covariance.

Conversely, consider a pure profile and suppress every route with a singleton eligible support. This reduced profile is admissible and has the same joint revenues. Covariance with $b=1$ restores the original profile by adding $s_e$ to each eligible segment's payoff. At the reduced profile, an eligible segment with $d_e=0$ is inessential. The allocation outside the eligible set is zero by the surcharge property when applicable. If $D=0$ this determines the allocation at the reduced profile. If $D>0$, proportionality for joint revenue requires $\psi_e=k d_e$ for every eligible segment with $d_e>0$. Efficiency gives $k=D/\sum_e d_e$. Adding back $s_e$ establishes uniqueness for each component, and separability determines the mixed allocation. The same argument with $N=E$ proves the pooled case. \hfill$\square$

\subsection{Proof of Theorem~\ref{thm:scs-recursive}}

Fix a connected graph $H$ and a root $i$. An admissible tree rooted at $i$ is obtained by choosing, for every component $K$ of $H-i$, a vertex $j\in J_i(K)$ and an admissible tree on $H[K]$ rooted at $j$. These choices are independent across components, and attaching each selected root to $i$ reconstructs the whole tree uniquely. Consequently,
\begin{equation}\label{eq:dfs-count}
n_i(H)=\prod_{K\in\mathcal C_H(V(H)\setminus\{i\})}
\left(\sum_{j\in J_i(K)}n_j(H[K])\right),
\end{equation}
with $n_i(H)=1$ on a singleton. Under the uniform distribution on trees rooted at $i$, the marginal probability of selecting root $j$ in component $K$ is $p_{ij}(H,K)$. Given that choice, the tree in $K$ is uniform among its admissible completions rooted at $j$.

For each admissible tree, the hierarchical allocations in \eqref{eq:scs-hierarchy} telescope to the worth of the entire graph. On a branch $K$ the hierarchical allocation is precisely the one calculated in its own rooted tree for $v|_K$. Averaging first conditional on the root therefore gives rooted efficiency and \eqref{eq:branch-consistency}. Averaging the conditional vectors with $p_i(H,w)$ gives each individual tree rooted at $i$ weight $w_i/\sum_z w_z n_z(H)$. This is \eqref{eq:scs-rule}.

For uniqueness, use induction on $|V(H)|$. On a singleton, rooted efficiency determines the sole payment. On a larger graph, each component of $H-i$ is smaller than $H$. By induction its conditional allocations are determined for every admissible branch root. Branch settlement consistency determines every entry of $x^i(H,v)$ except the root's entry. Rooted efficiency determines that remaining entry. Finally the initial root lottery determines $x(H,v;w)$ from the conditional vectors. \hfill$\square$

\subsection{Verification of the compensated sharing formula on a line}\label{app:scs}

For the unanimity game of a connected support $A$, consider an admissible tree with root $z$. If $z\in A$, removing $z$ from its descendant set leaves no child component that contains $A$, so $h_z^B(u_A)=1$. If $z\notin A$, connectedness places $A$ in one child component, and the contribution of $z$ is zero. Repeating the argument in that component gives exactly one route recipient. The coefficients in \eqref{eq:scs-hierarchy} are linear in $v^c$, so this argument applies to each route contribution separately.

For a line with $n$ segments and a route interval $[a,b]$, each possible root corresponds to a unique admissible tree. A root left of $a$ sends the revenue to $a$, one right of $b$ sends it to $b$, and a root in $[a,b]$ receives it itself. When $a<b$, uniform averaging therefore gives shares $a/n$ to $a$, $(n-b+1)/n$ to $b$, and $1/n$ to every interior segment. A singleton interval receives its whole revenue. Summing these terms over routes gives equation (5) of \citet{soto2026}.

\clearpage
\section{Logical independence of the axioms}\label{app:independence}

We test independence separately for Theorems~\ref{thm:additivity} and \ref{thm:abc}. For each axiom, we exhibit an allocation rule on the full model domain that satisfies every other axiom in its characterization and violates that axiom at an admissible profile. Some rules serve both characterizations. The examples below use a physical triangle with edges labelled $1,2,3$, mileage weights $(1,1,1)$, and surcharges $(1,1,0)$. The admissible route family includes the single-edge routes and the two-edge paths used below. Thus the examples apply on a cyclic network. All profiles cited in the examples can also be obtained from the tariff model by using a single vehicle class with mileage rate one and suppressing the other component when a pure profile is needed.

For each rule, unless stated otherwise, the mixed-profile allocation is the sum of the rule's allocations on the two pure counterparts. Every rule gives a nonnegative vector. When we describe a route-by-route rule, the contributions of all routes are added, and zero-revenue routes contribute zero. For any such rule that allocates each route's entire revenue within its physical segment set, both efficiency and sub-highway efficiency hold. Indeed, a positive-revenue route meeting a sub-highway lies wholly within it.

\subsection{Four tests shared by the two characterizations}

\paragraph{Separability.}
Use WPS on pure profiles and use route SES on profiles with positive revenue in both components. This rule satisfies every other axiom in both characterizations, because those axioms concern pure profiles and the rule agrees with WPS there. In the triangle, take one vehicle on the path with edges $\{1,3\}$. Its mileage revenue is 2 and its surcharge revenue is 1. Summing the two pure WPS allocations gives $(2,0,1)$. The mixed allocation under route SES is $(3/2,0,3/2)$. Hence separability fails.

\paragraph{Zero surcharge weight.}
Keep WPS in the mileage component. In the surcharge component, divide a positive route's entire surcharge equally among its zero-surcharge segments if it has any. For a route whose segments all have positive surcharges, use the WPS shares. This rule is route additive at fixed attributes and pays only segments on positive-revenue routes. Each route distributes its entire revenue within the route, so the two efficiency axioms hold. For any two eligible segments on all positive-revenue routes, the normalized contribution of a given route is either zero for both or $t_r^f/w^f(P_r^f)$ for both. This proves proportional treatment. For average balanced contributions, only routes containing both compared segments contribute to their two cross-effects. The same normalized terms, including the zero terms, appear in both cross-effects. Consequently the pairwise equalities hold, and summing them gives \eqref{eq:abc}. On the path $\{1,3\}$ with pure surcharge revenue 1, the allocation is $(0,0,1)$ even though segment 3 has surcharge weight zero.

\paragraph{Efficiency or sub-highway efficiency.}
Assign zero to every segment for every profile. The rule is separable and route additive. It satisfies the zero surcharge weight and inessential segment properties, and all proportionality and average balanced contributions equations hold with both sides zero. On a pure mileage profile with revenue 1 on the singleton route $\{1\}$, the total allocation is zero. This violates efficiency in Theorem~\ref{thm:additivity}. The singleton $\{1\}$ is also a sub-highway with revenue 1, so the same rule violates sub-highway efficiency in Theorem~\ref{thm:abc}.

\paragraph{Proportional treatment or average balanced contributions.}
Fix an ordering of the segments and pay each positive route's entire component revenue to its first eligible segment. The ordering is held fixed across profiles and starts with segment 1 in the example. The rule satisfies separability, zero surcharge weight, route additivity and the inessential segment property. Each route's payment stays within the route, which gives both efficiency and sub-highway efficiency. On a pure mileage profile with revenue 2 on the route $\{1,2\}$ and zero elsewhere, the allocation is $(2,0,0)$. Both segments are used on the only positive-revenue route and have weight one, so proportional treatment fails. Moreover, $\{1,2\}$ is a sub-highway. Cancelling either segment's routes produces the zero profile. For $i=1$ in \eqref{eq:abc}, the left side is 2 and the right side is 0. Thus the same rule violates weighted average balanced contributions in the second characterization.

\subsection{Two further tests for the additivity characterization}

\paragraph{Route additivity.}
For a pure profile $P$, let $U_c(P)$ be the union of the eligible segment sets of its positive-revenue routes, and let $T_c(P)=\sum_r t_r^c(P)$. Pay zero outside $U_c(P)$ and divide $T_c(P)$ across $U_c(P)$ in proportion to the component weights. If there are no positive-revenue routes, pay zero. This rule satisfies efficiency, the inessential segment property and zero surcharge weight. Eligible segments common to all positive-revenue routes belong to $U_c(P)$ and have the same normalized allocation. Separability follows from the definition of the mixed rule. Take two pure mileage profiles with positive revenue only on the single-edge routes $\{1\}$ and $\{2\}$, respectively, with revenues 1 and 3. The two separate allocations sum to $(1,3,0)$. On the sum of the profiles, the rule pays $(2,2,0)$, which violates route additivity.

\paragraph{Inessential segment property.}
In each component divide the total revenue over all of $E_c$ in proportion to its weights, regardless of route use. When $E_c$ is empty, the component revenue and the allocation are zero. This rule is separable and route additive at fixed weights, is efficient, and satisfies zero surcharge weight. Every eligible segment has the same allocation per unit of weight, so the proportionality requirement holds. On a pure mileage profile with revenue 1 only on the singleton route $\{1\}$, the rule pays $1/3$ to each segment. Segments 2 and 3 therefore receive positive amounts despite being inessential.

Both sets of axioms are therefore independent on the full domain. On a restricted domain, an axiom can become vacuous. For example, this happens to the zero surcharge weight property when every surcharge is positive.

\subsection{Independence of the route SPS axioms}

The six axioms in Theorem~\ref{thm:sps-axioms} are independent on the full componentwise domain. The four tests concerning pure rules other than zero surcharge weight also apply to the pooled characterization by taking full route supports. In the following tests, pure rules are combined by separability unless the test concerns separability itself. A fixed segment order is used wherever an allocation is assigned to a first segment.

\paragraph{Separability.}
Use componentwise SPS on pure profiles and route SES on mixed profiles. The pure axioms all hold. If a single route uses two equal-mileage-weight segments, one with surcharge weight zero, let it generate 2 in mileage and 1 in surcharge. The two pure SPS allocations sum to $(2,1)$ with the positive-surcharge segment first, whereas route SES on the mixed profile gives $(3/2,3/2)$.

\paragraph{Zero surcharge weight.}
Use ordinary SPS in the mileage component. In the surcharge component, consider the set of zero-surcharge segments occurring on positive routes with at least two eligible segments. If this set is empty, use SPS. Otherwise, keep $s$ and distribute half of $D$ among the eligible segments in proportion to $d$. Pay the remaining half equally to the members of that zero-surcharge set. Efficiency and the inessential segment property hold, and the eligible allocations are proportional to $d$ when $s=0$. The joint revenue and both joint allocations scale together, since the set of zero-surcharge recipients depends only on positive joint routes. If $b=0$ there is no joint revenue. The rule therefore satisfies weak toll covariance. A positive surcharge route containing two positive-surcharge segments and a third segment with surcharge zero violates zero surcharge weight.

\paragraph{Efficiency.}
Pay only the singleton revenues $s_e$. This preserves covariance and inessential and zero-weight treatment. All eligible payments are zero when $s=0$, so the cross-multiplied proportionality holds. The rule fails efficiency whenever $D>0$.

\paragraph{Inessential segment property.}
If there is an eligible segment with no positive joint revenue, allocate half of $D$ equally among all such segments and the other half among segments with $d_e>0$ in proportion to $d_e$. Keep $s$ in full. When no such eligible segment exists, use SPS. When $D=0$, pay $s$ alone. This rule is efficient and respects zero surcharge weight. It remains proportional among segments with $d_e>0$ when $s=0$. Its joint allocation scales with joint revenues for $b>0$ and vanishes for $b=0$, so covariance holds. A pure mileage profile with one positive two-segment route on a network with three eligible segments pays part of $D$ to the unused third segment.

\paragraph{Weak toll covariance.}
Use SPS on pure profiles with $s=0$ and WPS on other pure profiles. This respects efficiency, zero surcharge weight and the inessential segment property. Proportionality for joint revenue holds because the rule uses SPS whenever the axiom applies. For a pure mileage example, consider three equal-weight consecutive segments. Begin with unit revenues on the routes $\{1,2\}$ and $\{1,2,3\}$ and no singleton revenue. The allocation is $(4/5,4/5,2/5)$. Add one unit of singleton revenue at segment 3. The rule now uses WPS and gives $(5/6,5/6,4/3)$. The first two segments' allocations have changed despite an unchanged joint revenue profile. This violates covariance. For the pooled characterization the same rule uses route SES in place of WPS on profiles with positive singleton revenues, giving the same counterexample.

\paragraph{Proportionality for joint revenue.}
Keep $s$ and award all of $D$ to the first eligible segment with $d_e>0$ whenever $D>0$. The rule is efficient, excludes inessential and zero-weight segments, and preserves covariance since the joint-revenue participants are unchanged under positive scaling and the award vanishes when $b=0$. On a profile with positive revenue only on one two-segment eligible route, the two segments have equal positive $d_e$ but only one receives joint revenue.

\subsection{Independence of the recursive SCS axioms}

The following tests concern the joint family of conditional and unconditional allocations in Theorem~\ref{thm:scs-recursive}.

\paragraph{Rooted efficiency.}
Set both conditional and unconditional allocations to zero at every profile. The branch equation and initial averaging hold, but a positive singleton revenue is lost.

\paragraph{Branch settlement consistency.}
Conditional on $i$, give all available revenue to $i$, and use the required initial lottery for the unconditional allocation. Rooted efficiency and initial averaging hold. On two adjacent segments with a unit singleton revenue at segment 2, conditional settlement with root 1 gives nothing to segment 2. The singleton branch settlement after removal of segment 1 gives segment 2 one unit. This violates branch settlement consistency.

\paragraph{Consistency with the initial root lottery.}
Use the prescribed conditional tree averages, but take their unweighted mean over the possible roots for the unconditional rule. This preserves rooted efficiency and branch settlement consistency. For a two-segment graph with weights $(2,1)$ and unit revenue on the route using both segments, the alternative unconditional rule gives $(1/2,1/2)$. There is one admissible tree per root, so the prescribed lottery instead gives $(2/3,1/3)$.

\end{document}